%% file: template.tex
\documentclass[final,journal]{vgtc}              

\onlineid{8082}

\vgtccategory{Research}

\vgtcpapertype{Application/Design Study}

\title{XGraphRAG: Interactive Visual Analysis for Graph-based Retrieval-Augmented Generation}

\author{
  Ke Wang, Bo Pan, Yingchaojie Feng, Yuwei Wu, Jieyi Chen, Minfeng Zhu, and Wei Chen
}

\authorfooter{
  \item
      Ke Wang, Bo Pan, Yingchaojie Feng, Yuwei Wu, Jieyi Chen, and Wei Chen are with the State Key Lab of CAD\&CG, Zhejiang University.
      E-mail: \{sttotphd\,$|$\,bopan\,$|$\,fycj\,$|$\,22451008\,$|$\,chenjieyi\_juraws$|$\,chenvis\}@zju.edu.cn.
   
  \item
      Minfeng Zhu is with Zhejiang University.
      E-mail: minfeng\_zhu@zju.edu.cn.

  \item Minfeng Zhu is the corresponding author.
}

\abstract{
Graph-based Retrieval-Augmented Generation (RAG) has shown great capability in enhancing Large Language Model (LLM)’s answer with an external knowledge base. Compared to traditional RAG, it introduces a graph as an intermediate representation to capture better structured relational knowledge in the corpus, elevating the precision and comprehensiveness of generation results. However, developers usually face challenges in analyzing the effectiveness of GraphRAG on their dataset due to GraphRAG’s complex information processing pipeline and the overwhelming amount of LLM invocations involved during graph construction and query, which limits GraphRAG interpretability and accessibility. This research proposes a visual analysis framework that helps RAG developers identify critical recalls of GraphRAG and trace these recalls through the GraphRAG pipeline. Based on this framework, we develop XGraphRAG, a prototype system incorporating a set of interactive visualizations to facilitate users’ analysis process, boosting failure cases collection and improvement opportunities identification. Our evaluation demonstrates the effectiveness and usability of our approach. Our work is open-sourced and available at \url{https://github.com/Gk0Wk/XGraphRAG}.

}

\newcommand{\editt}[1]{\textcolor{black}{#1}}
\teaser{
  \centering
  \includegraphics[width=\linewidth, alt={A view of a city with buildings peeking out of the clouds.}]{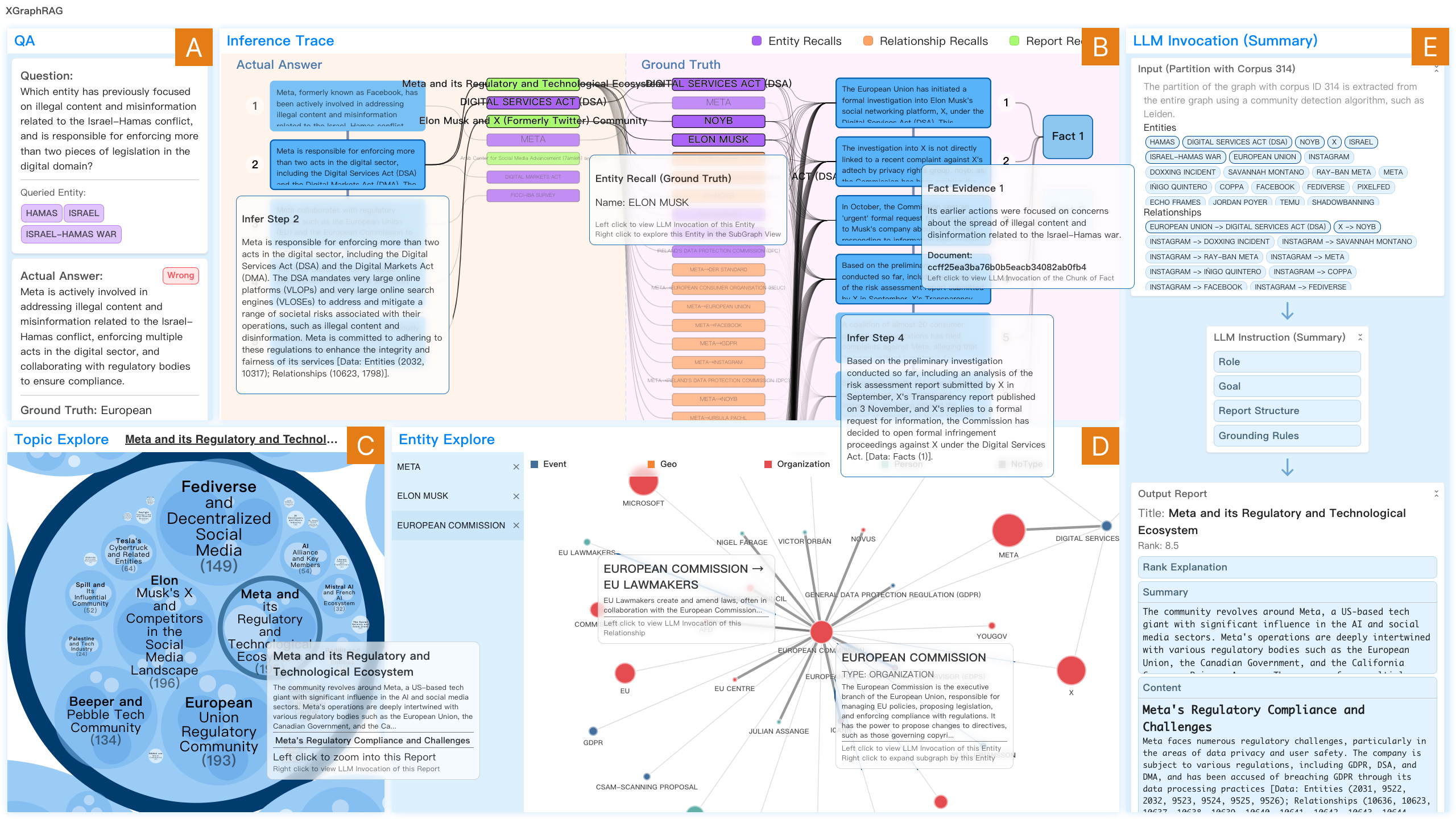}
  \caption{\editt{\textit{XGraphRAG} contains four views: (A\&B) The QA \& Inference Trace View allows users to locate suspicious recalls. (C) The Topic Explore View supports global relevance analysis on the graph for suspicious retrievals. (D) The Entity Explore View supports local relevance analysis on the graph for suspicious retrievals. (E) The LLM Invocation View presents details for each LLM invocation}.}
  \label{fig:teaser}
}

\graphicspath{{figs/}{figures/}{pictures/}{images/}{./}} 

\usepackage{tabu}                      
\usepackage{booktabs}                  
\usepackage{lipsum}                    
\usepackage{mwe}                       

\usepackage{amsmath}
\usepackage{amssymb}
\usepackage{mathtools}
\usepackage{graphicx}
\usepackage{CJKutf8}
\usepackage[normalem]{ulem} 
\usepackage[pagebackref,bookmarks]{hyperref} 
\begin{document}

\newcommand{\pb}[1]{\textcolor{black}{#1}}
\newcommand{\wk}[1]{\textcolor{black}{#1}}
\newcommand{\wkk}[1]{\textcolor{black}{#1}}
\newcommand{\edit}[1]{\textcolor{black}{#1}}


\maketitle

\input{chapters/1-introduction}
\input{chapters/2-related-work}
\input{chapters/3-overview}
\input{chapters/4-workflow}
\input{chapters/5-system}
\input{chapters/6-evaluation}
\input{chapters/7-discussion}
\input{chapters/8-conclusion}

\acknowledgments{
This work was supported by the Fundamental Research Funds for the Central Universities(No. 226-2024-00228), National Natural Science Foundation of China (No. 62132017, No. 62302435), ``Pioneer'' and ``Leading Goose'' R\&D Program of Zhejiang (No. 2024C01167), and Zhejiang Provincial Natural Science Foundation of China (No. LD24F020011). We want to thank Qianxing Wang, Haoran Jia, Hang Liu, and Haoyu Chen for their kind help. We also would like to thank the anonymous reviewers for their insightful comments.
}

\bibliographystyle{abbrv-doi-hyperref}

\bibliography{template}

\appendix 

\end{document}

%% file: chapters/1-introduction.tex
\definecolor{bdcolor}{RGB}{68,170,168} 
\newcommand{\overviewBox}[1]{%
\tikz[baseline=-0.5ex]{\node[rounded corners=0.2em, 
fill=white, 
draw=bdcolor, 
minimum size=0.9em, 
text=bdcolor, 
text centered,
inner sep=0pt,
line width=0.9pt,
font=\fontsize{9pt}{0.9em}\fontseries{ul}\selectfont](TsNode){#1}}}

\section{Introduction}

Large Language Models (LLMs) have rapidly gained prominence in recent years, demonstrating significant value and potential across various sectors such as industry\cite{tiro2023possibility,wang2023chatgpt}, healthcare\cite{thirunavukarasu2023large,abd2023large,sallam2023chatgpt}, science\cite{wardat2023chatgpt,agler2011waste}, and Finance\cite{wu2023bloomberggpt,yang2023fingpt}. Despite their unprecedented language comprehension and text generation capabilities, LLMs can face limitations when dealing with domain-specific knowledge, real-time updated information, and proprietary data, which are not in their pre-training corpus\cite{gao2023retrieval}. A promising paradigm for addressing this limitation is Retrieval-Augmented Generation (RAG), which integrates a retrieval component within the generation process to leverage the information contained in a customized large text corpus. This process not only enriches the contextual depth of the responses but also boosts their factual accuracy and specificity \cite{peng2024graph, fan2024survey}. Although RAG has achieved impressive results, it faces limitations in real-world scenarios that involve complex structures of relationships among different entities in the customized text corpus \cite{peng2024graph, procko2024graph}. To address this, recent works \cite{edge2024local,wu2024medical,guo2024lightrag} propose introducing graphs as an intermediate representation during the retrieval process (i.e. Graph-based RAG) to capture better and leverage the structured relational knowledge contained in the customized text corpus, bringing about new opportunities to elevate the precision and comprehensiveness of generation results in RAG applications.

However, despite the promising prospects of the Graph-based RAG (GraphRAG) paradigm, current RAG application developers often face challenges in analyzing the effectiveness of GraphRAG on their datasets. This reduces their confidence and motivation to deploy GraphRAG solutions in actual production. This difficulty arises from the two additional layers of complexity introduced by GraphRAG compared to traditional RAG: First, the information processing pipeline of a GraphRAG system is much longer and more complex than that of traditional RAG. Starting from raw text, the information goes through a series of intricate processes, including entity and relation extraction, graph construction, information aggregation and summarization, and parallel multi-stage retrievals. Users lack effective tools to track and analyze these processes. Second, the information processing in a GraphRAG system involves frequent and extensive calls to LLMs. It is difficult to understand the context and role of each LLM invocation within the graph during the analysis process.

As such, it is highly desirable to provide analysis support for GraphRAG developers to examine the GraphRAG process on their customized text corpus, enabling them to make informed design decisions and adjustments. To clarify the analysis targets, we survey recent GraphRAG-related papers and open-source projects to abstract a common pipeline for GraphRAG. For the sake of better understanding the underlying challenges in the analysis process, we conducted a formative study with 5 experienced RAG experts to have a grasp of the current analysis process, common analysis strategies, and the challenges they encounter when analyzing the GraphRAG process.

Based on the findings of the formative study, we propose a two-stage visual analysis framework for the process diagnosis of GraphRAG systems. In the first stage, it leverages the power of the LLM to provide automatic evaluation of model answers and identify the suspicious incorrect recalls that mislead the inference process. In the second stage, it allows users to trace the causes of the incorrect recalls through multi-facet relevance analysis on the graph. Based on this framework, we design \textit{XGraphRAG}, a visual analysis system consisting of four interrelated views. The QA View (Fig~\ref{fig:teaser}A) and the Inference Trace View (Fig~\ref{fig:teaser}B) serve as the springboard for exploration that helps users compare the discrepancies between the answer and ground truth to identify suspicious recalls. The Topic Explore View (Fig~\ref{fig:teaser}C) and the Entity Explore view (Fig~\ref{fig:teaser}D) allow users to conduct relevance analysis on the graph from different aspects. The LLM Invocation View (Fig~\ref{fig:teaser}E) adaptively displays structured details of the LLM invocation related to the intermediate output the user is interested in during the analysis process. We conduct a usage scenario demonstration and a user study to verify the effectiveness of our approach.

In summary, this paper makes the following contributions:
\begin{itemize}
\item We identify challenges in the analysis of the GraphRAG process and distill design requirements.
\item We propose a visual analysis framework that allows GraphRAG developers to examine the GraphRAG process efficiently and systematically.
\item We present \textit{XGraphRAG}, a prototype system that instantiates our framework, whose effectiveness is verified with a usage scenario demonstration and a user study.
\end{itemize}

%% file: chapters/2-related-work.tex
\section{Related Work}

Here, we review works in the field of GraphRAG and visual Analysis for LLM-based Applications, which are closely related to our study.

\subsection{GraphRAG}

GraphRAG is an emerging powerful retrieval-augmented generation paradigm that leverages knowledge graph as an intermediate representation to enable more precise and comprehensive retrieval for LLM generation, which has recently shown great potential in diverse application areas, such as medicine\cite{DALK}, E-commerce\cite{ecommerce}, intelligence analysis\cite{Ranade_2023}, and software engineering\cite{alhanahnah2024depsragagenticreasoningplanning}. While all GraphRAG approaches utilize graphs as a crucial intermediate knowledge representation, the underlying design space is vast and has attracted extensive research attention. For the graph construction phase, various indexing strategies (e.g. basic graph indexing\cite{gutiérrez2025hipporagneurobiologicallyinspiredlongterm}, text indexing\cite{li2024unioqaunifiedframeworkknowledge}, vector indexing\cite{he2024g}, and hybrid indexing\cite{sarmah2024hybridrag}) have been explored to facilitate downstream retrievals. For the graph retrieval phase, different types of recalls (e.g. nodes\cite{li2024graphneuralnetworkenhanced}, relationships\cite{li2024unioqaunifiedframeworkknowledge}, subgraphs\cite{edge2024local}, and paths\cite{ma2024think}) can be retrieved. Moreover, the retrieval paradigm can vary from efficient and cost-effective one-time retrieval \cite{he2024g, gutiérrez2025hipporagneurobiologicallyinspiredlongterm} to more sophisticated and adaptive iterative retrieval \cite{ma2024think, sun2023thinkongraph}.

While extensive work has focused on optimizing GraphRAG's process design, none has addressed how to assist developers in analyzing the underlying complex processes—a crucial aspect for practical GraphRAG applications. In this work, we propose a visual analytics framework that enables users to better understand and analyze GraphRAG processes, facilitating its effective deployment in real-world scenarios.

\subsection{Visual Analysis for LLM-based Applications}
Based on the interaction patterns of LLM invocations, existing visualization systems for analyzing LLM behavior can be categorized into two main directions: single-shot LLM invocation analysis and iterative LLM invocation analysis.

For single-shot LLM invocation analysis, researchers have developed various visualization systems to analyze individual LLM responses. PromptIDE \cite{strobelt2022interactive} enables users to experiment with prompt variations and visualize prompt performance for ad-hoc task adaptation. LLM Comparator \cite{kahng2024llm} facilitates side-by-side evaluation of LLM responses through interactive visual analytics, helping users understand performance differences between models. To analyze specific LLM capabilities, some work focused on specialized analysis tasks. For example, researchers proposed visualization techniques to analyze text style transfer \cite{BrathVisualizingLT}, revealing how LLMs apply different stylistic features in their responses. In terms of security analysis, JailbreakLens \cite{feng2024jailbreaklens} provides a visual analytics system for analyzing jailbreak attacks against LLMs, enabling users to evaluate model vulnerabilities and defensive capabilities.

For iterative LLM invocation analysis, existing work has explored how to visualize and analyze complex interaction patterns between multiple LLM calls. Sensecape \cite{suh2023sensecape} supports multilevel exploration and sensemaking with LLMs by enabling users to manage information complexity through hierarchical organization. ChainForge \cite{arawjo2024chainforge} provides a visual toolkit for prompt engineering and hypothesis testing across multiple LLM responses. Recent work has also focused on visually analyzing LLM-based agent systems. AgentLens \cite{lu2024agentlens} visualizes the dynamic evolution of agent behaviors in LLM-based autonomous systems, while AgentCoord \cite{pan2024agentcoordvisuallyexploringcoordination} helps users explore and design coordination strategies for LLM-based multi-agent collaboration. \cite{liu2023sprout} breaks down the programming tutorial creation process into actionable steps and visualizes the tree-of-thought exploration process.

Extending this research line, we aim to develop an interactive visual analysis framework for analyzing complex GraphRAG processes with diverse and intensive iterative LLM invocations.

%% file: chapters/3-overview.tex
\section{Formative Study}
We conduct a formative study to inform the design of our approach. First, to understand the core process of GraphRAG, we surveyed existing papers and high-impact open-source projects involving GraphRAG. Based on the survey, an abstraction of the common pipeline of GraphRAG is summarized in Section \ref{Common Architecture of GraphRAG}. Second, we conducted a formative interview with 5 RAG experts to identify the challenges encountered during the analysis of the GraphRAG process with existing tools in Section \ref{Common Architecture of GraphRAG}. Finally, we distill the four design requirements to improve the analysis process.

\subsection{Common Pipeline of GraphRAG}
\label{Common Architecture of GraphRAG}

\begin{figure}[htb]
    \centering
    \includegraphics[width=\linewidth]{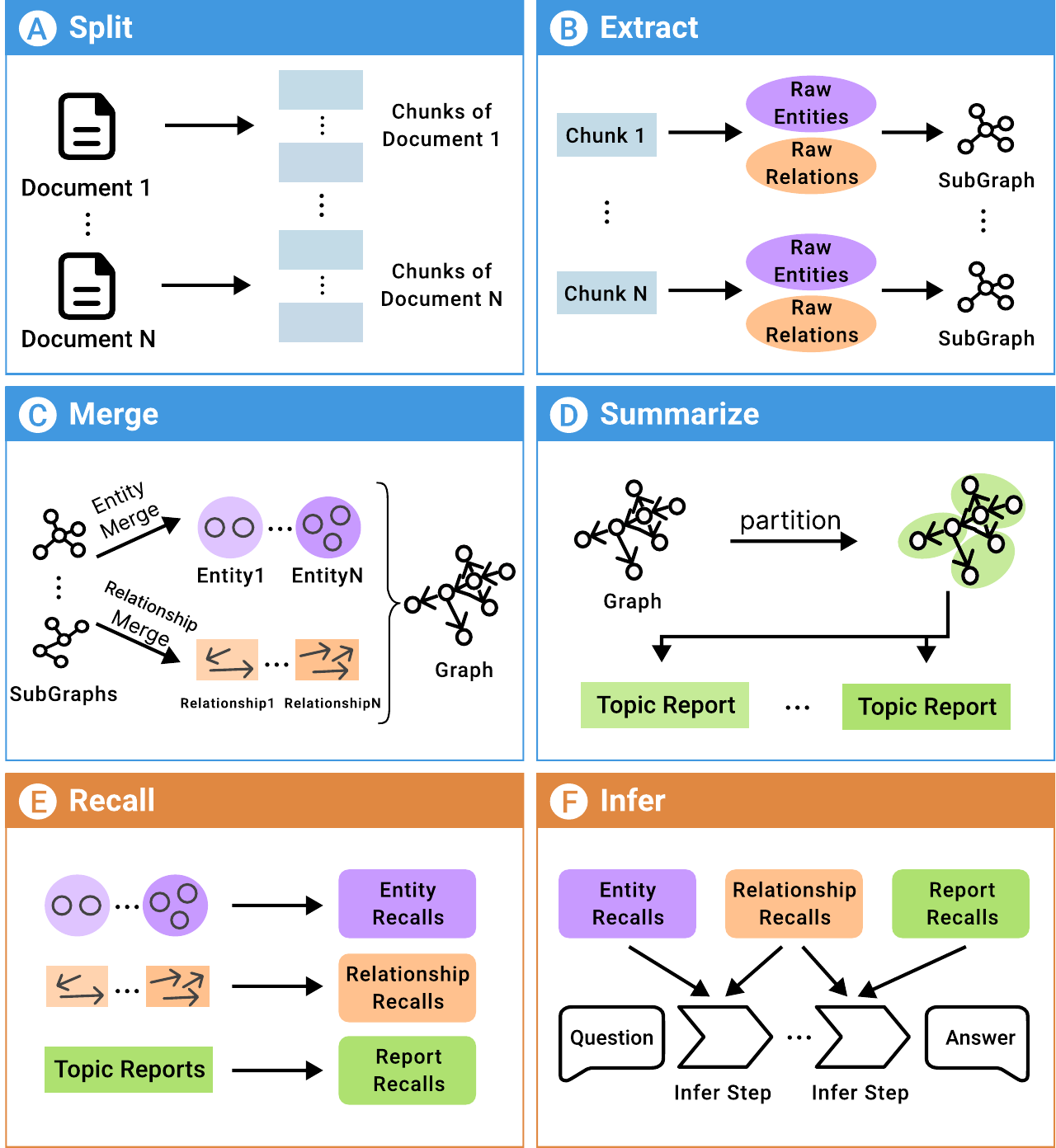} 
    \caption{Common Pipeline of GraphRAG, which is typically divided into an offline construction phase (a-d) and an online retrieval phase (e-f).} 
    \label{fig:common-architecture} 
  \vspace{-10pt}
    
\end{figure}

We survey recent GraphRAG-related papers and open-source projects to abstract the typical components of a GraphRAG pipeline. Given that GraphRAG is an emerging field with many works still awaiting peer review, we identified influential papers/projects in this area based on recommendations from RAG experts and search terms ``GraphRAG'', ``Graph Retrieval Augmented Generation'', ``Knowledge Graph RAG'', ``Graph-based RAG'', and ``Graph-augmented LLM''. Starting with several influential seeding works (e.g., GraphRAG\cite{edge2024local}, KAG\cite{liang2024kag}, LightRAG\cite{guo2024lightrag}), we iteratively expanded our corpus by analyzing both the references within these works and the works that cite them. Our final collection comprises 30 GraphRAG-related works (list included in our supplementary materials). Drawing from this corpus, we analyzed and summarized the essential data elements and data processing components shared by them. As shown in Fig~\ref{fig:common-architecture}, the GraphRAG pipeline is typically divided into two main components: an offline construction phase (a-d) and an online retrieval phase (e-f). We elaborate on the essential data process stages for each phase as follows.

\subsubsection{Construction Phase}
\label{Construction Phase}

\textbf{Split:} The initial step involves dividing the original documents into smaller, manageable \textbf{text chunks}. This segmentation is crucial as it sets the foundation for subsequent processing. The text chunks are designed to be coherent units of information, facilitating easier extraction and analysis in later stages.

\textbf{Extract:} For each text chunk, we extract \textbf{raw entities} and \textbf{raw relationships}, forming \textbf{subgraphs}. This involves identifying key elements within the text, such as names, dates, and specific terms, and understanding how these elements relate to each other. Each entity and relationship is enriched with additional metadata, such as \textbf{descriptions} and \textbf{entity types}, which provide context and clarity, enhancing the granularity of the subgraphs.

\textbf{Merge:} The next step is to combine these subgraphs into a comprehensive \textbf{graph}. This involves resolving conflicts where entities and relationships share names but differ in context. The merging process ensures that information is unified, with \textbf{entity} and \textbf{relationship} descriptions being consolidated. This step is critical for maintaining data integrity and ensuring that the graph accurately reflects the source material. Each \textbf{node} and \textbf{edge} in the graph is annotated with \textbf{chunk references} to the original text chunks from which they were derived, ensuring traceability.

\textbf{Summarize:} Once the graph is complete, it is partitioned based on the density of relationships among entities, resulting in distinct \textbf{topics}. This partitioning helps in organizing the graph into meaningful segments, each representing a coherent topic or theme. Within each topic, the entities and relationships are summarized to create concise \textbf{reports}. These summaries are constructed with clear references to the specific entities and relationships they are derived from, ensuring transparency. Depending on the complexity of the original corpus and the graph, this partitioning may occur at multiple hierarchical levels, providing varying degrees of abstraction and detail.

Throughout the construction phase, large language model invocations are extensively employed at different stages for natural language processing tasks. These models assist in tasks such as entity recognition, relationship extraction, and summarization, leveraging their advanced capabilities to enhance the accuracy and efficiency of the process.

\subsubsection{Retrieval Phase}

\textbf{Recall:} During the retrieval phase, user \textbf{queries} are utilized to extract relevant information from the constructed graph. The system identifies and retrieves various types of \textbf{recalls}, such as \textbf{entity recalls}, \textbf{relationship recalls}, and \textbf{report recalls}. This involves searching the graph for nodes and edges that are relevant to the user's query, ensuring that the most pertinent information is brought to the forefront.

\textbf{Infer:} The LLM then uses these recalls to perform step-by-step \textbf{inference}. Each inference step is explicitly linked to specific recalls, ensuring that the reasoning process is transparent and traceable. This iterative process allows the system to build upon retrieved information, synthesizing it to generate comprehensive answers. The inference culminates in producing a final \textbf{response} to the user's query, leveraging the depth and breadth of the graph's information.

This structured approach enhances the GraphRAG system's ability to manage and utilize complex relationships within data, supporting more effective retrieval and reasoning capabilities. By providing clear traceability and leveraging advanced language models, the system ensures both accuracy and transparency in its responses.

\subsection{Challenges of Analyzing the GraphRAG Process}
\textbf{Participants and procedure.} To understand the challenges users could encounter during the analysis of the GraphRAG process, we conducted a formative interview with 5 RAG experts. Three of them (E1-E3) are experienced RAG engineers who have first-hand experience with GraphRAG in their projects. The other two (E4-E5) are RAG researchers who have conducted at least one GraphRAG-related research project. In our interview, we first asked the participants about their current workflow for analyzing the GraphRAG process on their datasets and the challenges they encountered during this process. After that, we showed them how to use \textit{Kotaemon} \cite{kotaemon}, one of the most popular open-source projects that provide interactive analysis support for the GraphRAG process. We then asked the participants to try out \textit{Kotaemon} on their own in a think-aloud way (reporting the analysis target and the challenges they encountered in situ). We also collected participants' overall feedback at the end of the interview. Based on the interview, we summarize the challenges as follows.

\textbf{Lack of traceability across the complex information processing pipeline.} All of the participants mentioned the necessity and difficulty of tracing through the information processing pipeline of GraphRAG. To analyze the cause of the final or intermediate output, the participants need to "examine its upstream modules step by step" (E2). Nevertheless, this demand is not well-supported by existing tools. \textit{Kotaemon} \cite{kotaemon} provides some support for certain steps (e.g. users can click an entity to trace the chunk it is extracted from), but "lacks complete traceability from the final answer back to the raw document chunks" (E1).

\textbf{Lack of revealing of LLM invocation context on the graph.} The GraphRAG process typically involves hundreds or even thousands of LLM invocations to construct and query the graph. The participants mentioned that they often feel "overwhelmed and lost in the sea of LLM calls" (E2) and "hard to connect certain LLM invocation with its specific role on graph" (E5). Existing tools only provide exclusive or separate support for graph analysis for LLM invocation analysis. However, it is highly desirable to provide support for "analyzing LLM invocations on the graph" (E4). 

\textbf{Lack of support for multi-facet relevance analysis on the graph.} The key to the success of GraphRAG lies in whether it can utilize the graph structure to capture the relevance between the information in the customized corpus and the user's query. This relevance can be explicit local connectivity on the graph, or semantic connectivity based on global structure and specific techniques applied for graph summarization (as discussed in Section \ref{Construction Phase}). However, there is a lack of support for systematic analysis of different types of relevance on the graph. As E2 mentioned, "the key lies in finding why some relevance is successfully or unsuccessfully captured and exploited on graphs, which is a very complex and demanding analysis process"

\subsection{Design Requirement}

In response to the challenges identified in our formative study, we aim to design a visual analysis system to facilitate the analysis process for GraphRAG. The design requirements are distilled as follows:

\textbf{R1: Facilitate step-by-step trace for the generated answer.} To understand the final or intermediate output, users need to trace back to its upstream modules step by step. This traceability ensures that users can verify the accuracy of each step and identify the root cause of unexpected results. The system should provide flexible interaction and visualization to facilitate this tracing process.

\textbf{R2: Reveal LLM invocation context on the graph.} There is a large quantity of LLM invocations during the GraphRAG process. While those LLM invocations play a vital role in information extraction and processing during graph construction and query, it is hard for analyzers to link each invocation with its context on the graph during the analysis process. Therefore, it is highly desirable to provide support to help users understand the structure and context of those invocations.

\textbf{R3: Support global relevance analysis on the graph.} A significant advantage of GraphRAG over conventional RAG is its ability to use graphs to capture global semantic information. For example, if certain entities are not directly connected but are frequently mentioned in the context of certain topics, they might be grouped into the same community by GraphRAG, which is essential for answering questions that require a broader, more holistic understanding. Therefore, analysis support for the underlying process is important for understanding why GraphRAG succeeds or fails in different settings.

\textbf{R4: Support local relevance analysis on the graph.} While GraphRAG excels at capturing global semantic information, it also greatly enhances recall quality by leveraging local graph connectivity to capture local relevance. For instance, if a user's query pertains to a particular entity, understanding the direct connections and relationships of that entity can provide more precise and contextually relevant answers. Therefore, it is also important to provide support for a detailed, localized analysis of the graph.

%% file: chapters/4-workflow.tex
\section{Workflow}
Here, we elaborate on the workflow for analyzing the \textit{GraphRAG} process.
\label{section 4}
\subsection{Overview}

\begin{figure}[ht]
  \centering
  \includegraphics[width=\linewidth]{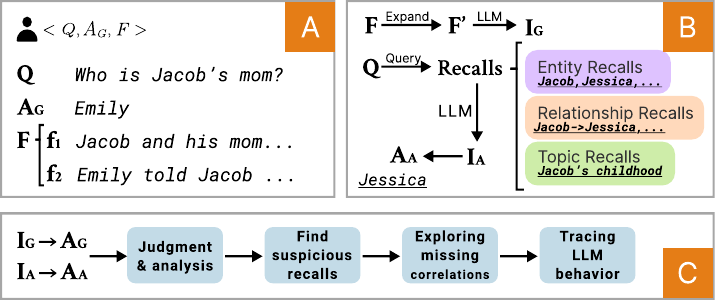}
  \caption{Overview of the workflow: The user first prepares a testing pair for evaluation (A). Subsequently, two recall-inference-answer chains are constructed for comparison, corresponding to the ground truth answer and actual answer respectively (B). Finally, users conduct a step-by-step analysis of the issue through a linear exploration process (C).}
  \label{fig:overview}
  \vspace{-10pt}
\end{figure}

To assist users in exploring graph construction issues, we divide the entire methodological process into two stages:

\textbf{Identifying Suspicious Retrievals:} Users conduct a question-and-answer evaluation using test data to preliminarily identify contradictions between actual answers and the ground truth. By analyzing the inference paths of both the actual answers and the ground truth, users can pinpoint problematic inference steps and identify suspicious recalls, understanding what the issues are and where they occur.

\textbf{Analyzing Suspicious Retrievals:} Starting with the identified suspicious recalls, users conduct either local or global analyses based on the type of recall. They trace reference chains to investigate the behavior of LLMs, uncover issues in graph construction, and provide evidence for optimizing graph quality, thereby understanding why the issues occur and how to address them.

\subsection{Stage 1: Identifying Suspicious Retrievals}

The GraphRAG system utilizes an initial document collection $D$ ($d_i \in D$) to construct a graph $\mathcal{G}$. During the testing phase, users query the constructed graph with test pairs $<Q, A_G, F>$ to retrieve actual answers $A_A$, where $Q$ is the question, $A_G$ is the Ground Truth Answer and $F$ is the set of evidence facts (Fig~\ref{fig:overview}A).
We evaluate both the correctness of the generated answers and their corresponding inference steps, guiding users to explore the unexpected results.
For \textbf{answer evaluation}, we employ an LLM to judge the answer $A_A$ as \textit{correct} or \textit{wrong} based on the Ground Truth $A_G$.
For \textbf{assessing inference steps}, we develop an LLM-assisted analysis framework \cite{wang2023vis+} to reconstruct the inference process for generated answers and ground truth and compare them to identify suspicious retrieval recalls (Fig~\ref{fig:overview}B).
We describe the details of the analysis framework as follows.

\subsubsection{Inferences Process Construction}

We construct the inference process as a \textit{Question-Recalls-Inferring-Answer} pipeline to facilitate comparative analysis.

\textbf{Actual Answer Inferences $I_A$.} When users submit a query, the GraphRAG system first retrieves a series of recalls. These are processed by the LLM to generate detailed inference steps ($I_{A_i} \in I_A$), each step indicating the recalls it relies on.

\textbf{Ground Truth Inferences $I_G$.} For the Ground Truth, each fact $F_i \in F$ from the test pair is contextually expanded to $F'_i \in F'$ to facilitate inferring, as the original facts may lack sufficient context. The expanded context test pair $<Q, A_G, F'>$ is then submitted to the LLM for reverse inferring, resulting in detailed inference steps $I_G$ based on $Q$ and $F'$, with each step specifying the facts utilized.

Each fact $F_i$ corresponds to several original document chunks. Entities and relationships related to these chunks are extracted to form a fact subgraph $\mathcal{G}_{F_i}$. For each inference step $I_{G_i}$, the relevant fact subgraphs are merged and filtered by the LLM to select entities and relationships pertinent to that step, forming the inference subgraph $\mathcal{G}_{G_i}$.

Finally, the entities and relationships in the inference subgraph $\mathcal{G}_{G_i}$ are used as recalls for each inference step $I_{G_i}$.

\subsubsection{Suspicious Recall Identification}

Based on the inference steps of Ground Truth $I_G$ and the Actual Answer $I_A$, users perform a comparative analysis to pinpoint problematic inference steps and identify the corresponding recalls for discrepancies. We flag two types of recalls as suspicious: \textbf{Missing Recalls}, which are essential to critical inference steps in the correct answer but absent from the actual inference steps, necessitating analysis to determine why they were not retrieved or used; and \textbf{Unexpected Recalls}, which do not belong to the critical inference steps in the correct answer but are erroneously included in the actual inference steps, requiring investigation to understand why they were mistakenly retrieved or utilized.

\subsection{Stage 2: Analyzing Suspicious Retrievals}

In the previous phase, users identified suspicious retrievals by analyzing problematic inference steps. To further investigate the causes of these issues in the GraphRAG system, users must dive into the internal structure of the system to explore the relationships between entities, relationships, and topic reports. This involves tracing the construction and retrieval processes to uncover the behavior of the LLM, continuing until the root cause of the issue is clarified (Fig~\ref{fig:overview}C).

\subsubsection{Exploring Missing Correlations}

The exploration of missing correlations is categorized into local and global investigations based on recall types, aiming to identify and address missing correlation information. \textbf{Local exploration} focuses on entity and relationship recalls, requiring users to examine their relationships with other entities and relationships. Specifically, these relationships may include connections with entities mentioned in the question, entities referenced in the correct answer, or entities involved in critical inference steps.

For \textbf{global exploration}, which centers on topic report recalls, users investigate their relationships with other reports. These relationships may include connections to topic reports that reference entities mentioned in the question or entities included in the correct answer. Through this structured approach, users can systematically trace missing correlations and enhance their understanding of the underlying retrieval process.

\subsubsection{Tracing LLM Behavior}

Once a recall with missing correlation information is identified, further tracing of the LLM's behavior is necessary to pinpoint where in the construction process this information was lost.

In the construction and retrieval stages of the GraphRAG system, each stage involves references to preceding content, with the LLM playing a crucial role. This forms the foundation for tracing and analyzing the behavior of the LLMs:

\begin{itemize}
    \item \textbf{Extract}: The LLM extracts multiple entities and relationships from chunks.
    \item \textbf{Merge}: The LLM consolidates same-named entities and relationships, referencing entity-chunks or relationship-chunks.
    \item \textbf{Summarize}: The LLM summarizes entities and relationships within a topic subgraph to create a topic report.
    \item \textbf{Infer}: The LLM uses query retrievals for reasoning, referencing inference step-entities and relationships.
\end{itemize}

To identify issues in different types of suspicious recalls, users undertake distinct tracing analyses. For entity and relationship recalls, the process involves analyzing the merging behavior and subsequently examining the extraction behavior. In the case of topic report recalls, a comprehensive investigation is conducted, focusing on the summarization, merging, and extraction behaviors.

%% file: chapters/5-system.tex
\section{System}
Here, we elucidate the design and usage of each view of \textit{XGraphRAG}. 
\subsection{QA View \& Inference Trace View}

\begin{figure}[ht]
  \centering
  \includegraphics[width=\linewidth]{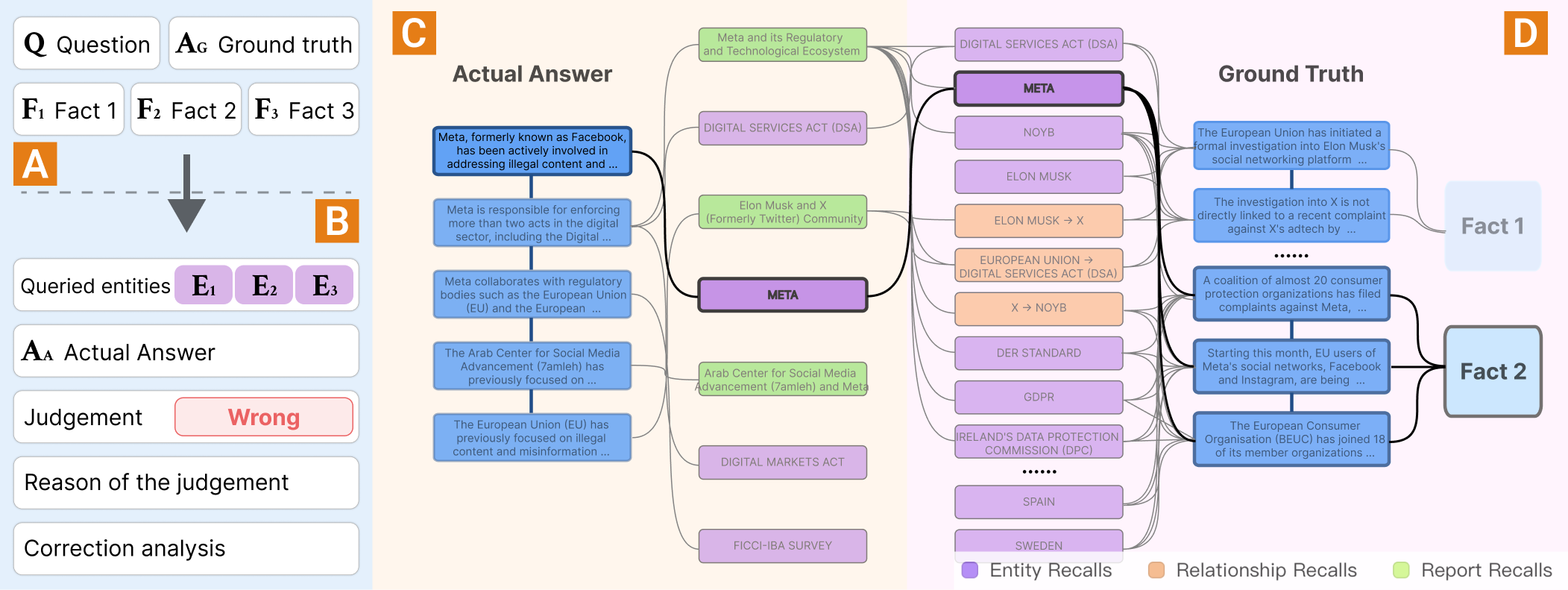}
  \caption{QA View \& Inference Trace View. Once the user inputs the test question in QA View (A), the system returns key information to support discrepancy analysis (B). A detailed comparison of the
inference steps for the Actual Answer (C) and the Ground Truth (D) are shown in the Inference Trace View.}
  \label{fig:qa_infer_trace_view}
  \vspace{-10pt}
\end{figure}

The QA (Query-Answer) View (Fig~\ref{fig:qa_infer_trace_view}A) and the Inference Trace View (Fig~\ref{fig:qa_infer_trace_view}B) serve as entry points for exploration, helping users identify discrepancies between Actual Answer (Fig~\ref{fig:qa_infer_trace_view}C) and Ground Truth  (Fig~\ref{fig:qa_infer_trace_view}D). They also help users locate problematic inference steps and identify suspicious recalls (\textbf{R1}). 

\subsubsection{Analysis of Question-Answer Discrepancies}

Once users input test questions in the form $<Q, A_G, F>$ and perform queries and analyses (Fig~\ref{fig:qa_infer_trace_view}A), the system initially presents key information to support discrepancy analysis in QA View (Fig~\ref{fig:qa_infer_trace_view}B). This includes the queried entities explicitly mentioned in the question, the actual answer generated by the GraphRAG system, and the corresponding Ground Truth. Furthermore, the system provides a relevance score that evaluates the alignment between the actual answer and the Ground Truth, offering a quantitative measure of their consistency.

To further support the analysis, justifications for the relevance assessment are provided along with explanations of any observed answer discrepancies. Users can view these justifications and explanations below the Ground Truth, enabling a detailed examination of the underlying causes of discrepancies.

This initial presentation provides users with a clear and structured overview, allowing them to gain a preliminary understanding of the differences between the actual answer and the Ground Truth. It serves as a foundation for a more in-depth analysis of the reasoning and retrieval processes.

\subsubsection{Analysis of Inference Steps}

Beneath the question-answer discrepancy analysis section lies the inference step analysis section, which provides a detailed illustration of the inference steps for both the Actual Answer and the Ground Truth. This section highlights the associated query recalls for each inference step, presenting a functionally symmetric view divided into two areas.

The left section is dedicated to analyzing the inference chain of the Actual Answer (Fig~\ref{fig:qa_infer_trace_view}C). It displays the sequence of inference steps and their associated query recalls, allowing users to trace the inferring that led to the Actual Answer. The right section focuses on the Ground Truth (Fig~\ref{fig:qa_infer_trace_view}D), presenting the reference inference steps alongside their associated query recalls and the factual basis for each step.

Users can systematically compare the inference steps of the Actual Answer on the left with the inference steps of the Ground Truth on the right. This side-by-side view enables users to identify discrepancies, pinpoint problematic inference steps, and analyze the differences in the inferring stage between the two.

To further aid analysis, query recalls associated with each inference step can be traced interactively. When users hover over an inference step, the related recalls and their connections are prominently highlighted, providing immediate context for the inferring stage. In the Ground Truth section, the interface also highlights the specific facts used in each inference step, helping users quickly locate the original basis for correct reasoning. This structured approach facilitates an in-depth examination of the inferring stage and its underlying assumptions.

\subsubsection{Analysis of Recall Usage}

Users can analyze the utilization of specific query recalls within the inferring stage by interacting with the system's interface. Hovering over a query recall provides detailed information on its role in the reasoning pipeline, highlighting the following information.

First, all inference steps that involve the selected query recall are displayed, offering a clear view of how the recall contributes to the reasoning process. Second, related recalls on the opposite side, including related inferences and recalls for both the Actual Answer and the Ground Truth, are presented. For topic recalls, this includes entities or relationships encompassed within the topic recall on the opposite side. For entity-relationship recalls, identical entity or relationship recalls on the opposite side are highlighted, along with included topic recalls on the opposite side.

Additionally, related inference steps for the recalls on the opposite side are displayed, where applicable. Finally, the source of facts for the recall is revealed, whether it pertains to the Ground Truth or to related recalls on the opposite side in the Actual Answer. 

This information enables users to comprehensively analyze the query recall’s usage in the inferring stage, compare the similarities and differences between the inferring steps on both sides and efficiently identify suspicious recalls.

\subsection{Topic Explore View}

\begin{figure}[ht]
  \centering
  \includegraphics[width=\linewidth]{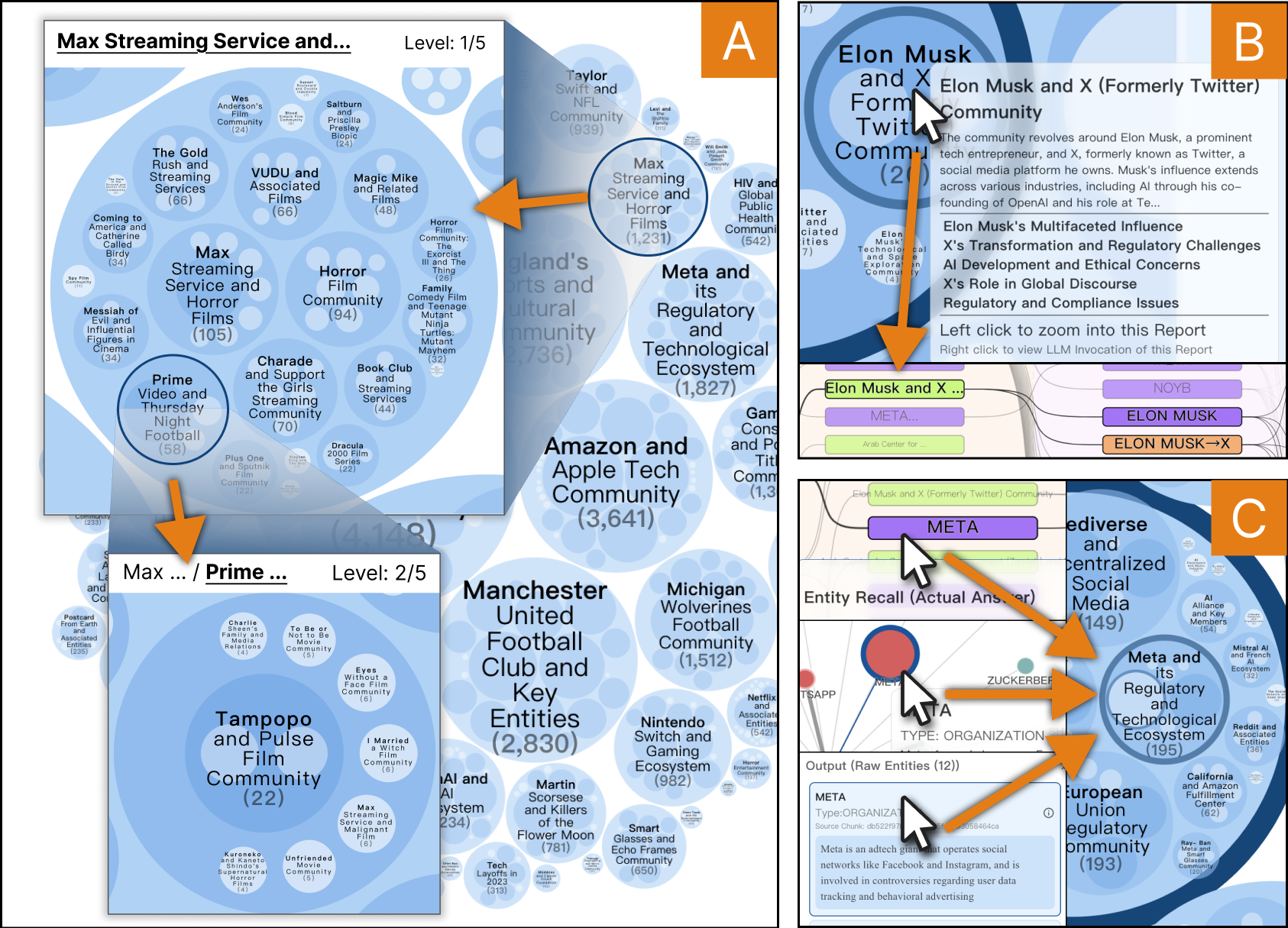}
  \caption{Topic  Explore View, which displays all topics with explicit nesting relationships in the graph (A). Whenever the mouse hovers over a topic, the corresponding topic-type recall in the Inference Trace View will be highlighted (B). Conversely, whenever the mouse hovers over any recall in the Inference Trace View, the topic related to it will be highlighted in the Topic Explore View (C).}
  \label{fig:topic_explore_view}
\end{figure}

The Topic Explore View (Fig~\ref{fig:teaser}C) utilizes the circle packing hierarchical visualization method to display all topics with explicit nesting relationships in the graph for global relevance exploration \textbf{(R3)}. Each topic is represented by a circle at a certain level, showing the topic's title and the number of entities it contains. The number of entities is encoded as the diameter of the circle to intuitively reflect the topic's global size. Although nested circles aren't as space-efficient as treemaps, the extra space helps to better display the hierarchical structure, helping users to understand the relationships between topics more intuitively. Within each topic circle, the next-level topic circles are drawn in a lighter color, providing an intuitive understanding of the composition of subtopics. Users can click on a topic to expand and display the next-level topics it contains or explore other topics at the same level, thereby progressively exploring the composition and structure of topics in the graph (Fig~\ref{fig:topic_explore_view}A).

The Topic Explore View also provides intuitive interactions to help trace the relationship between a topic and its related recalls: As shown in (Fig~\ref{fig:topic_explore_view}B), whenever the mouse hovers over a topic, the corresponding topic-type recall in the Inference Trace View will be highlighted. Conversely, as shown in Fig~\ref{fig:topic_explore_view}C), whenever the mouse hovers over any recall in the Inference Trace View, the topic related to it will be highlighted in the Topic Explore View. This facilitates quick exploration of global semantic relationships among different entities, relationships, and topics.



\subsection{Entity Explore View}

\begin{figure}[ht]
  \centering
  \includegraphics[width=\linewidth]{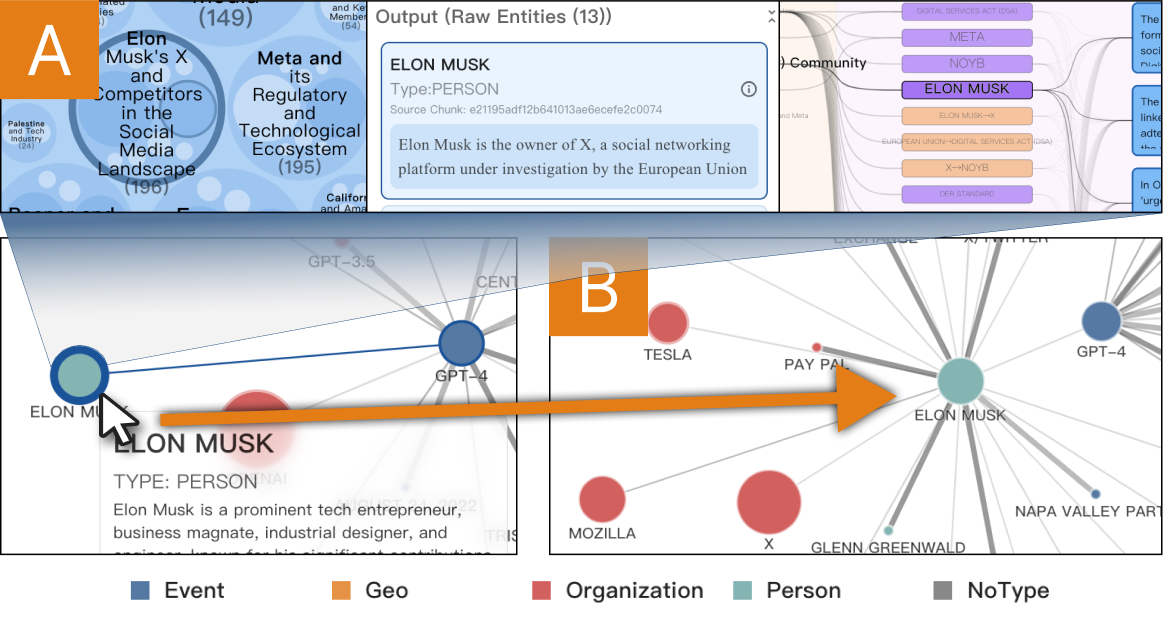}
  \caption{The Entity Explore View enables investigation of entity recalls through a node-link diagram. Node size encodes entity frequency, edge thickness shows topic distance, and node color indicates entity type. Interactive highlighting helps trace elements across different views (A). Users can expand the local graph by right-clicking nodes/edges to explore related entities and relationships (B).}
  \label{fig:entity_explore_view}
  \vspace{-10pt}
  
\end{figure}

The Entity Explore View (Fig~\ref{fig:teaser}D) assists users in investigating the local relevance of suspicious entity recalls with other entity recalls and relationship recalls \textbf{(R4)}. Whenever users identify a suspicious entity recall in the Inference Trace View, they can right-click to add it to the list in the Entity Explore View. Subsequently, users can utilize this view to explore other entities and relationships associated with the identified entity. We use a node-link diagram to visualize the local subgraph around this entity. Here the diameter of each node encodes the number of times it appears in different text chunks, intuitively reflecting the frequency and importance of the entity. The thickness of the edge encodes the distance of nodes in the nested topic tree, which also determines the strength of attraction in the force-directed layout. The node color encodes the types of entities. To ensure that each type has a distinctly different yet soft color, we use the HSL model, evenly sampling the hue values while maintaining moderate saturation and lightness. To help users better trace entities and relationships across views, whenever they hover over an entity point or relationship edge in the local graph, it will also be highlighted in other views (Fig~\ref{fig:entity_explore_view}A). If the user wants to explore more about a certain node/edge, they can right-click on it to add more entities and relationships near it (Fig~\ref{fig:entity_explore_view}B).


\subsection{LLM Invocation View}

\begin{figure}[ht]
  \centering
  \includegraphics[width=\linewidth]{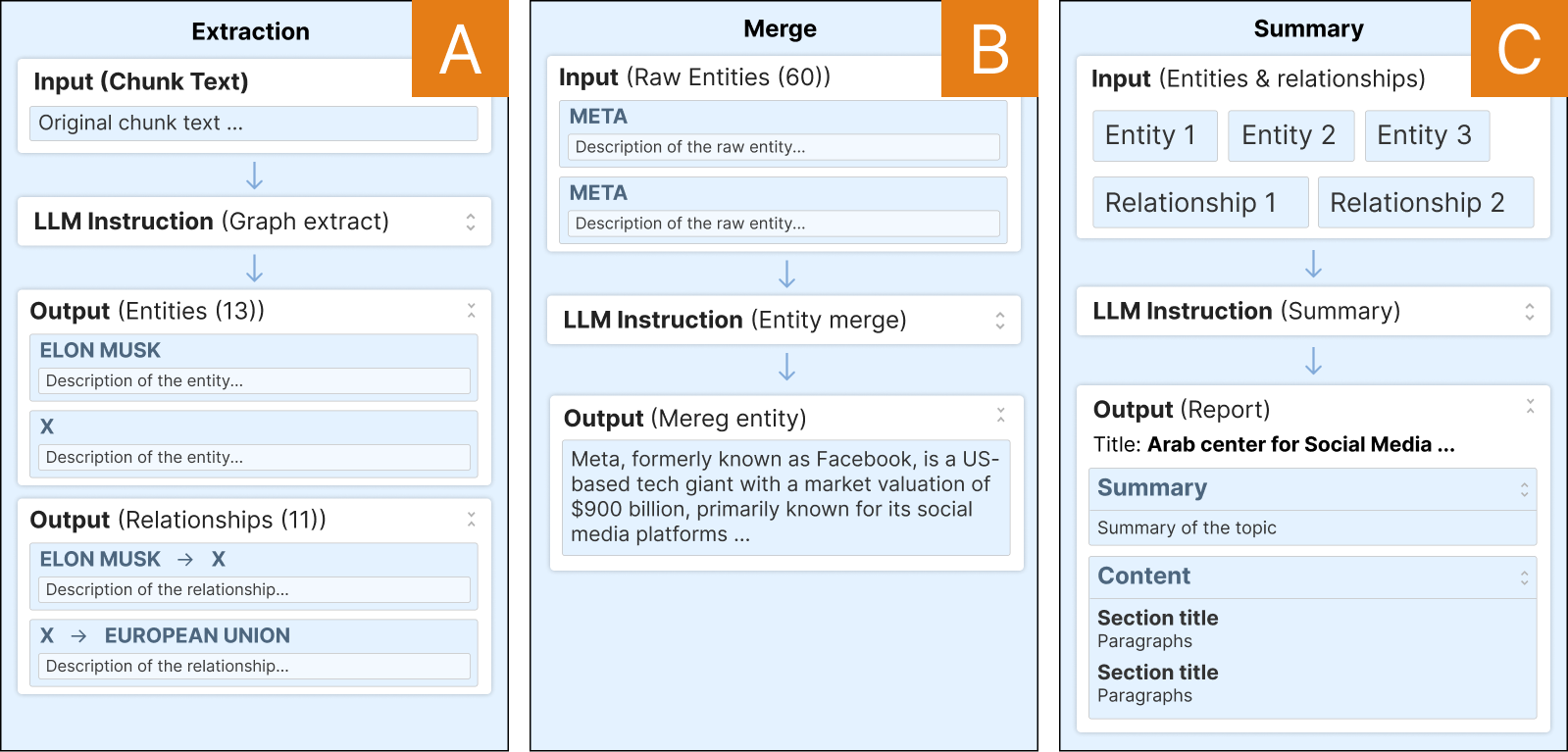}
  \caption{The LLM Invocation View has three variations, showing LLM behaviors during different stages of graph construction: extraction stage view displays source chunks and extraction details (A), merge stage view presents entity/relationship integration process (B), summary stage view reveals topic report generation with related entities and relationships (C).}
  \label{fig:llm_invocation_view}
  \vspace{-10pt}
\end{figure}

The LLM Invocation View (Fig~\ref{fig:teaser}E) reveals the behaviors of the LLM during various stages of graph construction in the GraphRAG system \textbf{(R2)}. This view has three variations for different stages, as elaborated below.

\subsubsection{Summary Stage LLM Analysis}

Whenever users click on any topic recall in any view, the related information of the topic and the behavioral details of the LLM in generating the topic report during the summary stage are displayed in the LLM Invocation View (see Fig~\ref{fig:llm_invocation_view}C). Users can investigate the entities and relationships used by the LLM in the summary process, the structure of the LLM's prompt words, and the topic report generated by the LLM. Since many entities and relationships may be used in topic summaries, this part can also link with other views for related information highlighting and filtering. For example, when the mouse hovers over an entity-relationship recall or fact in the Inference Trace View, or an entity point or relationship edge in the Entity Explore View, related entities and relationships are highlighted at the top of the list. This allows for quick confirmation of whether there is a relevant link between the part of the graph used for the topic and the object to be explored.

\subsubsection{Merge Stage LLM Analysis}

Whenever users click on any entity or relationship recall in any view, the related information of the entity/relationship and the behavior details of the LLM in processing the entity or relationship during the merge stage are displayed in the LLM
Invocation View (see Fig~\ref{fig:llm_invocation_view}B). Users can investigate the original entities or relationships used by the LLM in the merge process, the structure of the LLM's prompt words, and the merge results generated by the LLM. This information helps users understand how the LLM processes and integrates entity or relationship information from different sources and further traces the source of information for entities or relationships. Similarly, this view links with other views for hover-highlighting, enabling fluid tracing.

\subsubsection{Extraction Stage LLM Analysis}

Whenever users click on any fact or any entity or relationship during Merge Stage LLM Analysis, the source chunk information of the original entity or relationship and the behavior details of the LLM in the extraction stage are displayed in the LLM
Invocation View (see Fig~\ref{fig:llm_invocation_view}A). Then users can delve into how the LLM extracts entities and relationships from the original text using this information. This transparency helps users evaluate the reliability and accuracy of the information while also laying the foundation for further data quality improvement. Additionally, this view can help users trace the source of specific information, thereby better understanding and verifying the system's inference process.

%% file: chapters/6-evaluation.tex
\section{Evaluation}

Here, we verify the effectiveness of our method through a usage scenario
demonstration and a formal user study.

\subsection{Usage Scenario: Analysis on MultiHop-RAG Dataset}


This usage scenario illustrates how our system aids users in identifying and investigating issues related to graphs constructed by the GraphRAG system, thereby gaining insights into improving graph quality. Given the inherent complexity of the GraphRAG analysis process, we have also produced a demonstration video for this section to facilitate understanding.\footnote{Video URL: \url{https://gk0wk.github.io/XGraphRAG/}} The user employs the \textit{MultiHop-RAG} news dataset \cite{tang2024multihop} to construct a graph, intending to explore its quality and potential issues using our system.

The user inputs the query ``Which entity previously focused on illegal content and misinformation related to the Israel-Hamas conflict and is enforcing more than two pieces of legislation (act) in the digital domain?'' along with the Ground Truth Answer \textit{European Commission} and two key original text excerpts into the inference trace view for query and analysis.

Initially, the user observes that the entities \textit{Israel} and \textit{Hamas} were extracted from the query, ultimately leading to the answer \textit{Meta}. This answer is deemed \textit{wrong} by our system, which provides the reasoning for this judgment, highlighting two contradictions.

\begin{enumerate}
    \item \textit{Meta} is incorrectly identified as focusing on illegal content and misinformation about the \textit{Israel}-\textit{Hamas} conflict; this is actually the \textit{European Commission}'s role. (From Fact 1)
    \item \textit{Meta} is mistakenly thought to implement more than two digital acts, a responsibility that belongs to the \textit{European Commission}. (From Fact 2)
\end{enumerate}

It is noted that the query consists of two distinct parts: the first part tends to query the local association between explicit entities, while the second part aims to derive an answer based on a global summary topic. In this example, both parts result in contradictions, serving as an excellent illustration. Subsequently, the user investigates these two parts separately for local and global exploration.

\subsubsection{Exploration of Local Association Contradictions}

First, the reason why the segment ``Which entity previously focused on illegal content and misinformation related to the Israel-Hamas conflict'' failed to yield the correct answer should be analyzed. Issues with local queries can generally be attributed to a lack of entity or relationship information, so we will focus on exploring these two aspects.

By comparing inference steps, the user discovers that the steps related to \textit{Israel-Hamas} in the Actual Answer inference are steps 1, 3, 4, and 5. None of these steps utilized the \textit{European Commission} entity. However, in step 2 of the Ground Truth inference, entities such as \textit{Hamas} and \textit{European Commission} are used as recalls, indicating the presence of the \textit{European Commission} entity in the graph. Still, there is an omission when attempting to recall the \textit{European Commission} entity through \textit{Israel} and \textit{Hamas} entities.

Thus, the user seeks to explore whether there is a local association between the \textit{European Commission} and \textit{Israel} and \textit{Hamas}. In the Inference Trace View, the user right-clicks to recall the \textit{European Commission} entity and begins exploring its local relationships in the Entity Explore View. By hovering over \textit{Israel} and \textit{Hamas}, the user finds no highlighted edges or nodes in the Entity Explore View, indicating no direct relationship between \textit{Israel} and \textit{Hamas} with the \textit{European Commission}, resulting in a missing relationship on the graph.

Next, the user investigates the cause of the missing relationship. By left-clicking to recall the \textit{European Commission} entity in the Inference Trace View, the LLM Invocation View (for the merge stage) reveals the process of merging this entity from 11 raw entities. Hovering over ``Fact 1'' (a text segment describing the relationship between the \textit{European Commission} and \textit{Israel-Hamas}) in the Inference Trace View highlights one of the raw entities, indicating it originates from the chunk containing ``Fact 1''. However, the type and description of this raw entity are empty.

By clicking on ``Fact 1'', the LLM Invocation View (for the extraction stage) displays the chunks where the raw entity resides. It is found that the chunk does not extract any relationship with \textit{Israel} or \textit{Hamas}. Further reading of chunk text reveals that due to changes in subject and pronoun, the model fails to understand the relationship among the three, resulting in the failure of entity and relationship extraction and ultimately causing the loss of relational information.

\subsubsection{Exploration of Global Semantic Contradictions}

Next, the reason why the segment ``is enforcing more than two pieces of legislation (act) in the digital domain'' failed to yield the correct answer should be analyzed. Note that this answer originates from ``Fact 2'', where the \textit{European Commission} is responsible for enforcing the \textit{DMA} and \textit{DSA} digital acts on \textit{Meta}.

In the second step of the Actual Answer inference, information about \textit{Meta} fulfilling the \textit{DMA} and \textit{DSA} acts is indeed included, but the \textit{European Commission} is not mentioned. Further examination reveals that this inference step uses a topic report recall. Clicking on this topic recall, the LLM behavior analysis view shows the process of summarizing the topic into a report. Reading the report reveals descriptions of \textit{Meta}, the \textit{DSA}, and the \textit{DMA}, but no mention of the \textit{European Commission}. Further inspection of all entities and relationships used in this topic, hovering over ``Fact 2'', highlights entities and relationships related to ``Fact 2'' in the list, but the \textit{European Commission} is not found among them, indicating that the topic does not include the \textit{European Commission}, and there is no global semantic association between the \textit{European Commission} and the two acts.

Subsequently, the user further investigates the cause of semantic omission. In the previous analysis, the user suspected that the graph lacked the relationship from ``Fact 2'', which includes the \textit{European Commission} and the two acts. By left-clicking ``Fact 2'', the LLM Invocation View (for the extraction stage) displays the extraction process of the chunk. The user finds that although the chunk explicitly mentions the \textit{European Commission}'s responsibility for enforcing the two acts, the model fails to accurately capture this critical information during entity and relationship extraction. This discovery further confirms the user's previous suspicion that the LLM overlooked the extraction of the corresponding relationship.

\subsection{User Evaluation}
This study aims to evaluate the \textit{XGraphRAG} performance in analyzing \textit{GraphRAG} results, focusing on:

\begin{itemize}
\item \textbf{Efficiency}: The speed and simplification of analysis.
\item \textbf{Effectiveness}: The comprehensiveness and accuracy of the system's support.
\item \textbf{Usability}: The system's user-friendliness and ease of use.
\end{itemize}

It also compares \textit{XGraphRAG} with a baseline system for analysis support and user experience.



\subsubsection{Participants}
We invited 14 RAG engineers or experts as participants. These individuals have a thorough understanding of RAG principles but were not involved in the design requirements phase of this study, thereby enhancing the assessment's validity and the results' generalizability. Some have experience with naive RAG systems, while others are familiar with GraphRAG systems.

\subsubsection{Baseline Systems and Experimental Setup}

An additional baseline system was set up for comparative study \cite{feng2023promptmagician, feng2023xnli} alongside our system. Both systems are based on Microsoft's \textit{GraphRAG} \cite{edge2024local}, constructed using the \textit{MultiHop-RAG} dataset to build identical graphs, and utilize GPT-4 as the default LLM. The test dataset used in the experiments also comes from the \textit{MultiHop-RAG} dataset, including questions, answers, and graphs.

Baseline: We used \textit{Kotaemon}, a popular GraphRAG visualization interface system with high ratings on GitHub, as the baseline system. This system provides a visual question-and-answer interface for \textit{GraphRAG} \cite{edge2024local}, \textit{Nano-GraphRAG} \cite{gusye2024nanographrag}, \textit{LightRAG} \cite{guo2024lightrag}, and other systems. Users integrate GraphRAG for Q\&A through this system and display recalled entities, relationships, and topics referenced in the answers in the form of graph diagrams. It also shows various detailed information about recalls, such as explanations of entities. Users can click on entities and relationships in the graph to view the original chunk from which they originated.

Test Dataset: To test the ability of the GraphRAG system to handle both local and global questions, we improved the test questions of the test dataset. Firstly, we selected questions whose answers were names of people or organizations and ensured that these names had corresponding entities in the graph, with sufficient related text in the dataset. Then, we input these texts into the GPT-4 model to generate summary reports related to the entities. Subsequently, we verified the accuracy of the generated reports and posed global questions to them in a manner similar to the GraphRAG approach. Combining the original local questions, we formed the final test questions. Finally, to ensure the validity of the questions, we submitted these questions and related texts to three human experts for review. If the three experts could independently provide clear answers based on the texts and their answers were consistent, the question was deemed valid and could be used as a test case.

\subsubsection{Procedure and Tasks}

\textbf{Introduction (10 min):} We began by introducing the system to the participants, explaining the research motivation and methodology, and collecting basic information such as age and gender. With the participants' consent, we recorded and analyzed their usage behavior during subsequent tasks. Finally, we provided a detailed description of the features and characteristics of both the baseline system and the \textit{XGraphRAG} \cite{edge2024local} system, demonstrating their use through specific example scenarios.

\textbf{Task-based Analysis (60 min):} In this phase, participants interacted with both the baseline system and the \textit{XGraphRAG} \cite{edge2024local} system in a randomized order to eliminate any prior knowledge effects. The tasks were designed to assess the effectiveness and usability of each system, ensuring a fair level of complexity. We recorded the completion time and accuracy for each task to facilitate comparative analysis.

\textbf{Semi-structured Interview (30 min):} To further evaluate the interface design and usability of the systems, participants completed a questionnaire consisting of 6 items, using a five-point Likert scale to capture their perceptions and satisfaction. Participants rated each item from 1 (strongly disagree) to 5 (strongly agree). Concurrently, an open feedback session was conducted, allowing participants to explain the reasons behind their ratings and provide detailed insights into their user experience.

\subsubsection{Task Completion Analysis}

\begin{table}[h]
    \centering
    \makebox[\columnwidth][c]{\includegraphics[width=1.05\columnwidth]{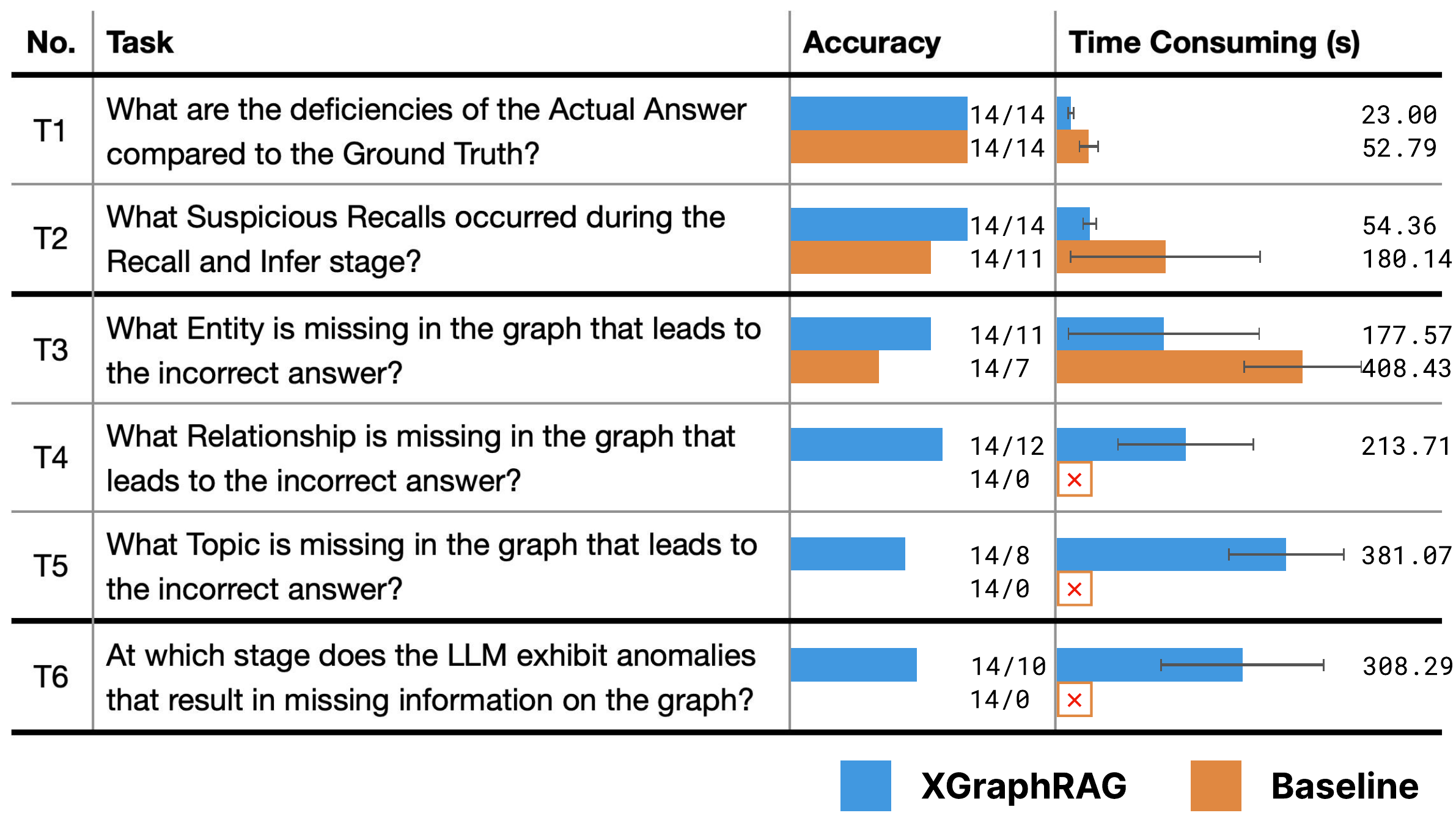}}
    \caption{Accuracy and time consumption comparison result between XGraphRAG and the baseline system.}
    \label{tab:resultcompare}
\end{table}

For each task (T1-T6, in Table\ref{tab:resultcompare}) and each participant, we randomly assigned a test case to explore. Before starting any task, users are provided with all necessary prerequisite information and acknowledge that they understand it. For example, ``It is already known that the question was answered incorrectly because the Google entity was not recalled; please try to find information related to the missing Google entity.'' (Example of T3) and ``It is known that the relationship between Taylor Swift and Travis Kelce is missing in the chunk where Fact2 is located; please explore the reason.'' (Example of T6). Participants need to arrive at an answer within the stipulated time to consider the task complete, such as ``The relationship of Epic filing a monopoly lawsuit against Google in Fact1 is missing.'' (Answer to the example of T3).

Although participants varied in accuracy and time consumption in tasks, using \textit{XGraphRAG} significantly improved task accuracy while reducing time consumption. Below, we analyze each type of task and provide the setup of task problems and objectives.

\textbf{Analysis of Suspicious Recalls (T1 - T2)}: This set of tasks focuses on the system's ability to identify contradictions and suspicious recalls in the results and their locations. Compared to the baseline system, \textit{XGraphRAG} performed more efficiently and accurately. Its graphical interface allows participants to quickly highlight contradictions in retrieval results in T1, while enhanced visualization tools help locate suspicious recalls in T2, thus reducing the need for extensive manual review and improving diagnostic accuracy ($p_1=0.00012,\ p_2=0.00012$).

\textbf{Analysis of Missing Information (T3 - T5)}: This set of tasks emphasizes the system's ability to help participants identify missing information in the graph. Compared to the baseline system, \textit{XGraphRAG} significantly improved time efficiency and success rate, especially in T4 and T5, where the baseline system's inability to analyze relationships and topic reports posed a considerable challenge for participants in analyzing global relevance. The visual encoding in \textit{XGraphRAG} helps participants effectively locate areas of missing nodes or relationships. This design provides clear visual cues, significantly reducing the effort required to locate and analyze incomplete parts of the graph ($p_3=0.0033,\ p_4=0.0022,\ p_5=0.012$).

\textbf{Analysis of LLM Behavior (T6)}: In this task, the baseline system lacks the functionality to present and analyze the behavior of the GraphRAG framework model, unable to assist participants in tracking factors leading to missing information in the graph. Using \textit{XGraphRAG}, participants can effectively identify and analyze LLM behavior at various stages of the GraphRAG process and accurately pinpoint issues related to missing information ($p_6=0.0051$).

Overall, the \textit{XGraphRAG} system excels in both accuracy and time efficiency in task completion. These findings indicate that the enhanced visualization tools and graphical reasoning capabilities of \textit{XGraphRAG} significantly improve participant efficiency and accuracy in complex analytical scenarios.

\subsection{Semi-structured Interview Analysis}

\begin{table}[h]
    \centering
    \makebox[\columnwidth][c]{\includegraphics[trim={10pt 100pt 265pt 10pt}, clip,width=1\columnwidth]{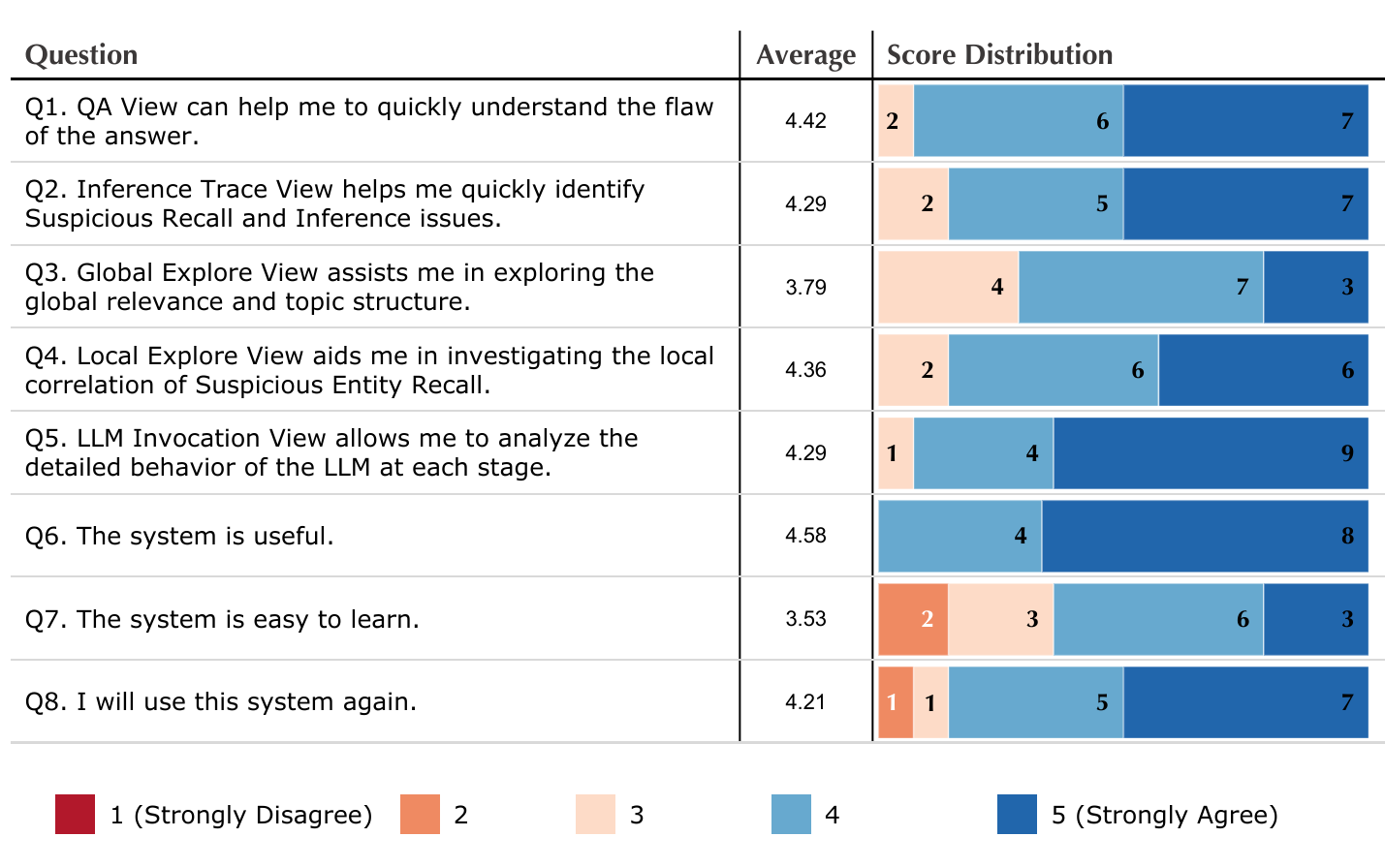}}
    \caption{The evaluation result for our system's effectiveness and usability.}
\vspace{-5pt}
\end{table}

We conducted a qualitative evaluation through semi-structured discussions to assess the system's usability and effectiveness. Participants were encouraged to provide detailed feedback on specific features, including their ability to identify reasoning process flaws, the clarity of visualizations, and the ease of interface interaction.

\textbf{Effectiveness:} All participants agreed that the Q\&A View effectively helped users quickly understand deficiencies in generated answers (Q1). P10 commented, ``The Q\&A View can automatically make judgments and provide reasons, along with three-point analysis suggestions, which greatly simplifies my thought process.'' P3 noted, ``In previous systems, I had to spend time reading the large outputs provided by the model, but now I can immediately see the key issues.''
Participants found the Inference Trace View particularly useful for identifying suspicious recalls and reasoning issues (Q2). P2 suggested providing an automatic translation feature for non-native speakers to accelerate understanding and analysis. P9 recommended integrating more context into the Inference Trace View, such as displaying additional information on mouse hover. P1 stated, ``Previously, I needed to use tools like neo4j to cross-system trace links for analysis, but now I can accomplish everything in one view, which gives me a sense of security.''
The Topic Explore View was highly praised for its assistance in exploring global correlations and corpus thematic structures (Q3). P8 emphasized, ``I've always wondered how to quickly understand what exactly a corpus is about, and this view has helped me; it's as important as a nautical chart.'' P11 stated, ``Even though the view is unnecessary for local problem analysis, I can quickly understand the distance between two unconnected entities by hovering over them, which is impressive and something I couldn't achieve with traditional graph tools.''
Participants also appreciated the Entity Explore View for its role in investigating local correlations of suspicious entity recalls (Q4). P13 pointed out, ``The view is simple and practical, allowing right-click expansion of entities for continuous exploration, and combined with the Topic Explore View, it enables targeted exploration.'' P1 suggested adding search filtering functionality to this view to handle large graphs.
The LLM Invocation View received positive feedback for its ability to analyze the detailed behavior of large language models at each stage (Q5). P5 stated, ``This is a missing link in GraphRAG and even RAG systems, and we need more interpretability work in LLM systems.'' P9 remarked, ``I want to understand the root cause of issues, and this view clearly shows each step, which is friendly to RAG developers; we no longer need to check lengthy run logs.''

\textbf{Usability:} Participants unanimously agreed that the system was useful for their tasks (Q6). P6 expressed, ``I previously tried tools like Kotaemon, but they only presented graph retrieval results, leaving a gap in presenting the complete chain from construction to retrieval reasoning, which made me reluctant to trust the model's answers. \textit{XGraphRAG} restored my confidence.''
Participants generally acknowledged the system's ease of use (Q7). We focused on feedback from participants who rated it 2 or 3. P2, P5, and others suggested adding a guided tutorial to help new users get started quickly. P7 mentioned, ``I hope there will be a mock example to guide users step-by-step through an analysis during their first use, which would enhance product usability.'' 
Most participants indicated they were likely to use the system again (Q8). Two low-scoring participants (Q8) acknowledged the system's functionality but had little need for diagnosing GraphRAG currently, so they were unsure about future use. P14 stated, ``XGraphRAG provides inspiration for designing subsequent Q\&A products in the industry and accelerates our development and testing processes.''

%% file: chapters/7-discussion.tex
\section{Discussion}

In this section, we discuss the system's generality, limitations, and directions for future improvements.

\textbf{System Generalizability:} Although we adopted Microsoft's \textit{GraphRAG} \cite{edge2024local} as an integration tool in our study, all designs are based on our proposed general architecture. This architecture is compatible with other mainstream GraphRAG systems. For instance, the six stages of our architecture are reflected in all systems, where recall, entity, and relationship are present and serve as the core functionalities. While topic reports may have different meanings and expressions in different systems, we regard them as a type of global summary recall. Other systems may have additional types of recall, leading to extra stages, but this does not conflict with the GraphRAG architecture.




\textbf{Enhancing temporal analysis.} While our system demonstrates satisfactory performance, dynamic changes in entity relationships or topics within a graph, such as the relationship between A and B evolving from friends to enemies, require further investigation. Therefore, it is essential to develop advanced temporal analysis tools to better guide users in exploring the temporal evolution of graphs.

\textbf{Expanding recall capabilities.} Although we currently support entity, relationship, and topic views in mainstream GraphRAG systems with promising results, the development of future GraphRAG systems will necessitate supporting more types of recall to accommodate a wider range of scenarios. This expansion represents another critical direction for improving system capabilities.
















%% file: chapters/8-conclusion.tex
\section{Conclusion}

This paper presents \textit{XGraphRAG}, a visual analysis system designed to enhance the understanding and optimization of GraphRAG processes. By addressing key challenges such as traceability, LLM invocation context, and relevance analysis, \textit{XGraphRAG} offers a comprehensive framework that empowers developers to systematically examine and refine GraphRAG systems. Through our formative study and user evaluations, we demonstrated the system's effectiveness in identifying and resolving graph quality issues, thus improving the precision and comprehensiveness of generated results. The integration of interactive visualizations and structured exploration tools not only facilitates more informed decision-making but also advances the field of visual analytics in complex data environments. Future work will focus on expanding the system's capabilities to support dynamic, real-time datasets and enhance user guidance for open-ended exploration.